\documentclass{vow2008}

\title{Multi-wavelength identification of high-energy sources.}
\author{R. P. Mignani}
\affil{Mullard Space Science Laboratory, University College London, Holmbury St. Mary, Dorking, Surrey, RH5 6NT, UK}

\usepackage{graphicx}
\begin{document}


\maketitle

\begin{abstract}
The nature of most of the $\sim$ 300 high-energy $\gamma$-ray sources discovered by the  {\em EGRET} instrument aboard the {\em Gamma-ray Observatory} ({\em GRO}) between 1991 and 1999 is one of the greatest enigmas in high-energy astrophysics.  While about half of the extragalactic sources have been optically identified with Active Galactic Nuclei (AGN), only a meagre 10\% of the galactic sources have a reliable identification. Interestingly, a few of them (7 in total) have been identified with Isolated Neutron Stars (INSs) thanks to the coincidence with known pulsars and to the discovery of $\gamma$-ray pulsations at the expected period.  The low success rate in the identification of $\gamma$-ray sources in the crowded regions of the galactic plane has mainly to be ascribed to the local crowding of potential optical counterparts and to the large $\gamma$-ray error boxes (of the order of  one degree in radius) which prevented a straightforward optical identification. Indeed, a multi-wavelength identification strategy, based on a systematic coverage of the $\gamma$-ray error boxes, has been the only do-able approach.\\
  The situation is now greatly improving thanks to the observations performed by the {\em Fermi} Gamma-ray Space Telescope (launched in June 2008) which, thanks to the {\em LAT} instrument, provides a factor of  50 improvement in sensitivity and a factor of 10 improvement in positional accuracy.  However, while the smaller error boxes will make the multi-wavelength follow-ups easier, the larger sensitivity will enormously increase the number of detected $\gamma$-ray sources,  requiring an even larger effort in the multi-wavelength follow-ups. This effort can not be obviously sustained by targeted observations only and it would greatly benefit from multi-wavelength data and advanced data products available world wide through the science data centres and interfaced by the Virtual Observatory (VO) tools. In this contribution, I outline the science case, the multi-wavelength observation synergies, and the requirements for both the the science data centres and the VO.  
\end{abstract}

\section{Introduction}

Launched in 1991, the {\em Gamma-ray Observatory} ({\em GRO}) was the second of the NASA's Great Observatories after the {\em Hubble Space Telescope} ({\em HST}).  In less than eight years of operational lifetime, the {\em GRO} performed the deepest mapping ever of the $\gamma$-ray sky at energies above 50 MeV with the {\em EGRET} instrument. In particular, it discovered almost 300 new $\gamma$-ray sources (Hartman et al. 1999), more or less equally distributed between extragalactic ($|b| > 10^{\circ}$) and galactic ($|b| < 10^{\circ}$), corresponding to almost a factor of 100 improvement with respect to the results of the previous high-energy $\gamma$-ray observatory, the ESA's {\em COS-B} satellite.  {\em GRO} observations clarified the nature of the sources detected at high-latitude,  half of which have been identified with different classes of Active Galactic Nuclei (AGN), thus joining the quasar 3C 273, the first extragalactic $\gamma$-ray source discovered and identified by {\em COS-B}.  However, only 10\% of the galactic $\gamma$-ray sources were identified during and after the {\em GRO} mission (see Caraveo 2008, for a recent review). This meagre success rate is mainly to be ascribed to the large error box of the $\gamma$-ray sources (of the order of one degree in radius) which prevented the straightforward optical identification in the crowded regions of the galactic plane. \\
The nature of the unidentified galactic $\gamma$-ray sources thus remained a mystery after the end of the {\em GRO} mission and it has been subject to many speculations.
  Interestingly, most of the few identified galactic sources turned out to be Isolated Neutron Stars (INSs), unambiguously identified as such by the discovery of $\gamma$-ray pulsations following the positional coincidence with a radio  pulsar. It is thus natural to assume that many other unidentified galactic $\gamma$-ray sources might be INSs. However, it most cases the poor $\gamma$-ray statistics and the lack of a positional coincidence  with a known radio pulsar hampered the search for a periodicity which would automatically certify these sources as INSs. Indeed, some INSs (if not the majority) may actually turn out to be radio-silent, like Geminga, the first INS of this class to be discovered (Bignami \& Caraveo 1996 and references therein), which would obviously depreave the astronomer from a radio reference period for the timing analysis. Since the large $\gamma$-ray positional uncertain heavily affects the reconstruction of a coherent pulsed signal, this is almost certainly unrecognized in a "blind" periodicity search, which hampered many identification efforts in the past years. Of course, all known classes of galactic X-ray emitters, e.g. X-ray binaries, microquasars, magnetars, supernova remnants (SNRs), are also viable counterparts to some $\gamma$-ray sources. In addition, it can not be excluded that some  $\gamma$-ray sources might be indeed associated with entirely new classes of astrophysical objects, whose existence passed unnoticed so far.  Apart from INSs which can be identified via the detection of $\gamma$-ray pulsations, by itself not always feasible, no other class of galactic high-energy sources features an evident identification signature. Thus, in most cases $\gamma$-ray source identification studies can be pursued only through a systematic multi-wavelength approach. 

\section{The multi-wavelength identification approach}

The multi-wavelength identification approach, successfully tested in the case of Geminga (see Bignami \& Caraveo 1996 for a summary), passes through a  sequence of "top-down" steps, i.e. consisting of observations at increasing wavelengths and angular resolution. The first step is  the systematic mapping of the $\gamma$-ray error boxes through X-ray observations to pinpoint possible X-ray counterparts to the $\gamma$-ray source, under the assumption that a $\gamma$-ray source should also be detectable at its "nearest neighbour" wavelengths. The second step is the mapping of the detected X-ray source positions with optical/IR observations to search for potential counterparts. This yields to the X-ray source identification and classification and thus to the selection of the best X-ray source  candidate counterpart to the $\gamma$-ray source,  which ultimately yields to its identification. \\
The X-ray source identification process is complex and  requires different types of information. From the X-ray side, classification evidence comes from the X-ray spectrum, from the derived hydrogen column density  $N_H$ which, together with the X-ray coordinates, gives an indication on the source distance and thus on whether it is galactic or extragalactic, from the X-ray flux {\em hardness ratio (HR)}, which is characteristic of the X-ray source class, and, of course, from long/short-term X-ray variability and periodicity. Another parameter typical of the X-ray source class is the ratio between the X-ray and optical flux $F_X/F_{opt}$ (see, e.g. Voges et al. 1999). On the optical side, classification evidence comes from the spectral energy distribution (SED) of the optical counterparts, which requires either long slit spectroscopy (for the brightest objects) or a photometry coverage in at least 5 passbands for a reliable classification, from optical variability, from an optical proper motion, which obviously suggests a galactic source, and from the object morphology (stellar or extended). \\
 Of course, barring the large amount of observing time involved to map  a full $\gamma$-ray error box in both X-rays  ($\sim 150$ ks) and in the optical ($\sim 80$ hours for five passbands), the identification of a hundred (or more) X-rays sources by assembling the collected X-ray/optical information through a human supervised decision tree is  obviously a time consuming process.  
Because of that, the multi-wavelength identification approach  has been applied only for a few, well-selected, cases  like, e.g., in La Palombara et al. (2006) where coordinated X-ray/optical observations with {\em XMM-Newton} and the ESO/MPG {\em Wide Field Imager} ({\em WFI}) allowed the identification of likely INS counterparts for the $\gamma$-ray sources 3EG\, J0616$-$3310 and J1249$-$8330, capitalizing on their better positional accuracy ($\approx$ 0.5 degrees) and on the $33\arcmin \times 33\arcmin$ {\em WFI} field of view which well matches that of the {\em EPIC} instrument aboard {\em XMM-Newton}. Similar coordinated X-ray/optical observations yielded to the identification of a handful more $\gamma$-ray sources (see La Palombara et al., 2006, and references therein). As it can be expected, a systematic multi-wavelength follow-up for, e.g.  all the $\sim 80$ unidentified galactic {\em GRO} $\gamma$-ray sources with error box smaller than one degree has never been tried since it would have required at least 10 Ms of {\em XMM-Newton} time to obtain a reasonably deep X-ray coverage  of all error boxes, and the equivalent of one year of observations with a 2.5m-class survey telescope. Such  an amount of observing time, pipeline data reduction load, and data interpretation overheads is only affordable to large, world-wide collaborations.

\section{The AGILE and FERMI Gamma-ray missions}

Perspectives for $\gamma$-ray astronomy became brighter with the launch  of the {\em AGILE} and {\em GLAST}, now {\em Fermi} Gamma-ray Space Telescope, satellites. {\em AGILE}, launched on April 2007, launched by the Italian Space Agency (ASI) mounts a main instruments for high-energy $\gamma$-rays: the {\em Gamma-ray Imaging Detector} ({\em GRID})  which covers the 30 MeV-50 GeV spectral range with a sensitivity equal (or slightly better) than that of the {\em GRO/EGRET}. The {\em GRID} instrument has a field of view of 3 steradiants (about 1/4 of the sky) and a positional accuracy better than 0.3 degrees, far better than that of {\em EGRET}. Furthermore, {\em AGILE} is equipped with the {\em Super-AGILE} detector which works in the hard X-ray range (10-40 keV), with a smaller field of view (0.8 steradiants) but a positional accuracy better than 3$\arcmin$, which makes it extremely useful for prompt or parallel coverage of the $\gamma$-ray error boxes. For a detailed description of the {\em AGILE} mission see Tavani et al. (2008).  {\em Fermi}\footnote{http://fermi.gsfc.nasa.gov/}, launched by the NASA on June 2008, mounts the {\em Large Area Telescope} ({\em LAT}) which has a factor of 50 higher sensitivity with respect to the {\em GRO/EGRET}. Compared to the {\em AGILE/GRID}, the {\em LAT} instrument has a larger spectral coverage (20 MeV-300 GeV), a comparable field of view (2.5 steradiants), but a positional accuracy better than 0.1 degrees. In the first year of operation {\em Fermi} will perform a deep all-sky scan (3 hours per single scan) which should lead to the detection of about 10000 $\gamma$-ray sources with a significance better than $5 \sigma$.  \\
Among the early results, {\em AGILE} observations yielded to the identification of the {\em GRO} source 3EG\, J2021+3716 with the radio pulsar PSR\, J2021+3651 (Halpern et al. 2008). More spectacularly,  {\em Fermi} has identified a new $\gamma$-ray pulsar (Abdo et al. 2008) with the unidentified source 3EG\, J0010+7309 in the CTA 1 SNR.  This was possible only thanks to the  unprecedented positional accuracy of the {\em LAT} instrument, crucial to reduce pulsar timing problems, which has allowed to detect pulsations from INSs even without the aid of a reference radio period.  At the time of writing this contribution, {\em Fermi} has already discovered that 11 more, so far unidentified,  $\gamma$-ray sources are indeed $\gamma$-ray pulsars which, like that in the CTA 1 SNR, are radio-silent. This is a further evidence that many unidentified galactic $\gamma$-ray sources are likely radio-silent INSs, as put forward a long ago since the identification of Geminga in the early 1990s. However, it has to be reminded that while many unidentified $\gamma$-ray sources can be straightly identified as INSs by {\em Fermi} through the detection of $\gamma$-ray pulsations, some of them might be either too faint for a periodicity search or might not be INSs but other classes of astronomical objects for which no timing signature is expected.  Thus, multi-wavelength observations will still play a crucial role in the identification of the newly detected $\gamma$-ray sources.

\section{Exploiting the multi-wavelength archives and the VO}

The multi-wavelength identification of serendipitous $\gamma$-ray sources detected by {\em Fermi} in the first year all-sky scan is a major project of the {\em LAT} collaboration which is coordinating dedicated follow-ups both in the X-rays, e.g. with {\em XMM} and {\em Swift}, and in the optical, e.g. with the {\em Gemini} and the {\em VLT}. While the improved positional accuracy of the {\em LAT} instrument ($<$0.1 degrees) results in much smaller fields to be covered by multi-wavelength mapping, thus reducing the follow-up observation load for a given source, its unprecedented sensitivity results in many more detected sources, thus increasing the number of fields to be mapped. This means that any large scale $\gamma$-ray source identification campaign will require an even more huge multi-wavelength follow-up work which can not be accomplished only through a programme of targeted observations. \\
In this respect, an important resource is represented by the huge amount of data available in the multi-wavelength archives maintained by world-wide science data centres, which help to reduce the observation loads. However, the level of the  data available in different archives is not even. In the X-rays, post operational archives (POAs) are available for the {\em ROSAT} and {\em ASCA} missions and incremental archives are available for {\em Chandra}, {\em XMM-Newton}, and {\em Swift}, in all cases providing the user with fully processed and calibrated data sets.  In addition, in several cases archival X-ray data sets come with associated object catalogues with basic information on the X-ray source properties (position, flux, spectrum, extension, etc.).  This is the case of the {\em ROSAT} catalogues\footnote{ttp://www.mpe.mpg.de/xray/wave/rosat/catalogue/index.php} of the all-sky survey and of the pointed observations.  Recently, X-ray catalogues have been produced both for {\em Chandra}, an early release based on the first three years of operations (Romano et al. 2008; Romano et al., these proceedings), and for {\em XMM-Newton} (Watson et al. 2009). Indeed, the latter (a.k.a. {\em 2 XMM}), is the largest X-ray catalogue ever produced, with a total of $\approx$ 200\,000 unique sources (i.e., accounting for multiple detections) over more than 500 square degrees. At the same time, the compilation of the first {\em Swift} X-ray source catalogue is under way. In the future, this already huge database will be increased by observations performed by the {\em eROSITA} X-ray telescope\footnote{http://www.mpe.mpg.de/projects.html\#erosita}, to fly in 2011 as part of the payload of the french/russian {\em Spectrum-X-Gamma} satellite. With a field of view of $1^{\circ} \times 1^{\circ}$ and an improved sensitivity to higher energies  {\em eROSITA} will perform the first X-ray all sky survey in the 0.1-12 keV band, yielding at least a factor of 10 increase in limiting flux with respect to the {\em ROSAT} all sky survey. 
While the wealth of  databases of X-ray sources is certainly a major advantage, the vast majority of them is still unclassified, which represents the major show stopper in the multi-wavelength $\gamma$-ray source identification flow. There are several reasons for that. \\
The first reason is that accurate X-ray source identification surveys, based on optical spectroscopy follow-ups, are affordable for a few selected fields only and are limited to the brightest field objects. On the other hand, identification surveys on larger scales, and down to fainter flux limits, can be performed based on multi-band imaging photometry and the associated object catalogues.  
In this case, VO tools would be important to perform more customised object matching between X-ray and optical catalogues using, e.g. different matching radii depending on the quoted positional accuracy of the X-ray and optical sources.  This would decrease the number of mismatches and it would easy the identification process. 
It is clear that an identification approach based on catalogue matching can be handled only  if automatic X-ray source classification tools are available. Assembling such tools, based on a human unsupervised decision tree algorithm  is obviously not easy.   A very interesting prototype tool for  X-ray source classification was {\em ClassX} (McGlynn et al. 2002) which, however, was originally fine-tuned for {\em ROSAT} sources only.  A more advanced tool has clearly to be versatile enough to adapt to X-ray sources detected by different satellites and instruments since some classification parameters, like the computed X-ray flux and the {\em hardness ratio}, critically depend on the assumed energy band.  Future X-ray source classification tools could take advantage of already available optical classification tools developed by the VO scientists, like e.g. {\em VOSpec} (Baines et al. these proceedings) which classifies an object based on its observed SED and the comparison with model spectral libraries.  Further advancements for such tools would be to allow SED fits to account for variable interstellar extinction and to account for different colour transformations between the photometry of different object catalogues, derived from observations taken with different instrument and filters, calibrated with respect to different photometric systems, and sometimes expressed in different units (e.g. Vega magnitudes or AB magnitudes). This is crucial if, for instance, one would like to derive a consistent SED from  the photometry of the {\em SDSS}, the {\em GSC-2}, and {\em 2MASS} catalogues.  \\
The second reason is that most of the currently available public optical databases  do not provide an adequate support for a systematic X-ray source identification work. Indeed, so far most of these works have been carried out using object lists matched from the {\em GSC-2}, {\em USNO-B1.0}, and the {\em SDSS} in the optical, which have a limiting magnitude of $B\sim 22$, and from {\em 2MASS} in the near-infrared (NIR), which has a limiting magnitude of $Ks \sim 15$.  Of course, this represents a severe limitation since the deeper flux limits now reached by the available X-ray catalogues requires similarly deeper optical catalogues to homogeneously sample the $F_{X}/F_{opt}$ parameter space. Thus, much deeper optical/NIR object catalogues are needed. 
An important step forward will be made thanks to the new optical/NIR sky survey which will be performed with dedicated 4m-class survey telescopes like, in the southern hemisphere, the ESO {\em VST} and the ESO/UK {\em VISTA}. Survey data will be fully reduced/calibrated by data reduction pipelines and multi-band object catalogues will be produced, ready to be used.    For the {\em XMM-Newton} sources, a valid support will come for the recently released serendipitous catalogue of objects detected by its {\em Optical Monitor (OM)} telescope, which provide multi-band optical and near ultra-violet (NUV) photometry down to $B\sim 24.5$ for the longest pointings (see Still et al. these proceedings).  Of course, much deeper serendipitous surveys can be built from observations performed with 8m-class telescopes, like the {\em VLT}, the {\em Gemini}, and with the {\em HST}.  Although such surveys would be mostly incomplete in terms of colour coverage, they would certainly represent a valuable support for at least some X-ray identification works. Unfortunately, only rarely the science data centres provide advanced data products, e.g. fully reduced and calibrated frames, image stacks, and  object catalogues which would be needed for this project. This is the case, for instance, for the {\em HST} for which science-ready, on-the-fly re-calibrated, imaging data and image stacks are available through the STScI, ST-ECF, and CADC web sites. Next to come it is probably a multi-instrument {\em HST} object catalogue.  The availability of advanced data products would certainly be an important commitment for other science data centres. For, e.g. the  {\em VLT} the possibilities of accessing advanced data products through the ESO science archive are  in perspective very good since each imaging instrument is supported by dedicated data reduction pipelines  and, in most cases, by object detection tools.

\section{Summary}

So far, only about half of the $\sim$ 300 $\gamma$-ray sources discovered by the NASA {\em GRO} satellite (1991-1999) have been identified, and the rate reduced to less than 10\% for the galactic ones.  Thus, the nature of the galactic $\gamma$-ray sources  has been so far one of the greatest enigmas in high-energy astrophysics. Solving this enigma is one of the major science goals in the next decade.  This goal will be pursued starting from observations performed by the new generation of $\gamma$-ray observatories: {\em AGILE} (launched in 2007) and especially {\em Fermi} (launched in 2008).  Apart from a better characterisation of sources already detected, many of which will be straightly identified thanks to the improved statistics, {\em Fermi} observations will yield to the detection of thousands more $\gamma$-ray sources which will await for identification.  In many cases, this will be  feasible only through a coordinated,  systematic multi-wavelength approach. On a large scale, this is a huge task which is unsustainable for targeted observations. However, it will certainly benefit  of the enormous amount of data available in world wide archives. Of course, a better scientific exploitation of this data requires the help of both the various science data centres to provide the much needed advanced data products, like reduced/calibrated data and associated object catalogues, and of the VO to provide data analysis tools to make  the matching of multi-wavelength catalogues easier and to provide automated object classification tools. 

\section*{Acknowledgments}
RPM acknowledges STFC for a Rolling Grant

\end{document}